\pdfoutput=1

\documentclass[11pt]{article}

\usepackage[final]{acl}

\usepackage{times}
\usepackage{latexsym}
\usepackage[T1]{fontenc}
\usepackage[utf8]{inputenc}
\usepackage{microtype}
\usepackage{inconsolata}

\usepackage{amsmath,amssymb,amsfonts}
\usepackage{graphicx}
\usepackage{textcomp}
\usepackage{xcolor}
\usepackage{booktabs} % For professional looking tables
\usepackage{multirow}
\usepackage{listings} % For code snippets
\usepackage{hyperref} % For hyperlinks
\usepackage{subcaption} % For subfigures
\usepackage{natbib} % For Harvard-style citations

\definecolor{codegreen}{rgb}{0,0.6,0}
\definecolor{codegray}{rgb}{0.5,0.5,0.5}
\definecolor{codepurple}{rgb}{0.58,0,0.82}
\definecolor{backcolour}{rgb}{0.95,0.95,0.92}

\lstdefinestyle{mystyle}{
    backgroundcolor=\color{backcolour},   
    commentstyle=\color{codegreen},
    keywordstyle=\color{magenta},
    numberstyle=\tiny\color{codegray},
    stringstyle=\color{codepurple},
    basicstyle=\footnotesize\ttfamily,
    breakatwhitespace=false,         
    breaklines=true,                 
    captionpos=b,                    
    keepspaces=true,                 
    numbers=left,                    
    numbersep=5pt,                   
    showspaces=false,                
    showstringspaces=false,
    showtabs=false,                  
    tabsize=2
}
\title{Can MLLMs Detect Phishing? A Comprehensive Security Benchmark Suite Focusing on Dynamic Threats and Multimodal Evaluation in Academic Environments}

\author{Jingzhuo Zhou \\
  School of Computer Science and Engineering \\
  UNSW Sydney, Sydney, Australia \\
  \texttt{z5549606@ad.unsw.edu.au} \\}

\begin{document}
\maketitle

\begin{abstract}
The rapid proliferation of Multimodal Large Language Models (MLLMs) has introduced unprecedented security challenges, particularly in phishing detection within academic environments. Academic institutions and researchers are high-value targets, facing dynamic, multilingual, and context-dependent threats that leverage research backgrounds, academic collaborations, and personal information to craft highly tailored attacks. Existing security benchmarks largely rely on datasets that do not incorporate specific academic background information, making them inadequate for capturing the evolving attack patterns and human-centric vulnerability factors specific to academia. To address this gap, we present AdapT-Bench, a unified methodological framework and benchmark suite for systematically evaluating MLLM defense capabilities against dynamic phishing attacks in academic settings. AdapT-Bench is built on four pillars: (1) A multi-dimensional evaluation methodology covering dynamic user states (trusting, anxious, stressed) and multi-modal inputs (text and vision). (2) Our experiments in the specific context of phishing emails revealed a "quality filtering paradox": we found that as we moved from a raw, randomly generated dataset to one created with specific attack templates, and finally to a most comprehensively filtered dataset, the more "perfect" an attack became, the progressively easier it was for AI systems to detect. (3) A cross-lingual and cross-model comparative protocol assessing state-of-the-art systems from OpenAI, Google, DeepSeek AI, and xAI under both Chinese and English prompts. (4) A specialized assessment for malicious jailbreak injections that simulate real-world scenarios. We release the dataset and evaluation pipeline to foster reproducible research on secure and trustworthy AI systems in high-risk academic environments.
\end{abstract}

\section{Introduction}

Multimodal Large Language Models (MLLMs) are rapidly proliferating and being integrated into a wide array of applications, from content generation to critical infrastructure control. However, their advanced capabilities are a double-edged sword, creating new attack surfaces for malicious actors. The attack paradigm has also evolved: on one hand, adversaries can leverage MLLMs to generate highly convincing phishing emails, spread misinformation, or craft complex social engineering attacks at an unprecedented scale; on the other hand, when these models are used as security assistants, the models themselves become direct targets of attack, for example, by using malicious prompt injection to hijack their judgment logic. Therefore, conducting comprehensive and in-depth evaluations of the security and robustness of these models has become critically important.

However, existing security benchmarks exhibit severe deviations in simulating real-world scenarios \citep{greshake2023prompt}. First, they generally lack specific academic background information, making them unable to simulate spear-phishing attacks that leverage personal information to build trust and a sense of urgency. Second, and more critically, they completely ignore the core variable of the 'human,' failing to consider the user's actual emotional state. The essence of a phishing attack is a psychological one \citep{khadka2024survey}, designed to exploit human emotions such as anxiety, trust, or stress to impair judgment. A benchmark that disregards the user's emotional state cannot measure a model's true protective capabilities for a user at their most vulnerable.

More importantly, this evaluation approach, being detached from reality, also fails to genuinely test the core capabilities of MLLMs. Using generic, context-free scam emails to test an MLLM is akin to giving a PhD in mathematics an elementary arithmetic test. This type of evaluation not only severely underestimates the complexity of real-world composite attacks that leverage personal information and psychological states, but also completely fails to assess the model's intrinsic robustness against direct adversarial attacks (such as prompt injection \citep{greshake2023prompt}). This leaves us with little knowledge of the defensive potential of MLLMs against complex, real-world threats.

Furthermore, a critical yet often overlooked real-world scenario is that many scholars, particularly within the vast Chinese-speaking academic community, habitually use Chinese as their primary language for interacting with models, even when processing English content. This gives rise to a unique cross-lingual security evaluation scenario: a user employs a Chinese prompt (e.g., "Help me check if this English email is a scam?") to ask an MLLM to assess an entirely English scam email. Our preliminary research and hypothesis suggest that in such cross-lingual scenarios, the model's accuracy in scam detection shows a noticeable decline compared to when purely English prompts are used for the same task.

This decline can likely be attributed to several factors. First, the cross-lingual task itself increases the model's cognitive load and potential for information loss, making it less sensitive to the subtle cues within the English text that indicate a scam, such as unnatural phrasing or specific luring tactics. Second, the training data and safety alignment of current mainstream MLLMs are predominantly English-centric, and their built-in safety mechanisms may not be as effectively triggered by Chinese prompts. Therefore, evaluating the security performance of MLLMs in this realistic, cross-lingual interaction model is crucial for protecting non-native English-speaking user groups, and it addresses a blind spot common to most current security benchmarks.

To address the aforementioned gaps, we present AdapT-Bench, a comprehensive and extensible benchmark designed to evaluate MLLM security in a more realistic and holistic manner, with a specific focus on academic environments. Within our framework, we systematically evaluate the security of MLLMs under various simulated emotional states (neutral, trusting, suspicious, anxious, stressed, confident) and across different modalities (text- and image-based attacks). To avoid the potential social impact of using real personal information, we constructed a synthetic scholar profile database. Based on this database, our framework generates multiple datasets that incorporate different attack strategies and varying levels of complexity, including the Kimi 1000, Strategies Batch 1, and Filtered Clean datasets.

A comparative analysis of the datasets revealed a distinct trend in false negative rates (FNR). The Kimi 1000 dataset ($n=1,000$) yielded the highest FNR. In contrast, the Strategies Batch 1 dataset ($n=1,000$), derived from four specific attack templates, demonstrated a reduced FNR, while the Filtered Clean dataset ($n=1,054$), having undergone the most comprehensive filtering, exhibited the lowest FNR. This observation leads to a counter-intuitive conclusion: attacks that are meticulously crafted and ostensibly sophisticated are, in fact, more readily identifiable by AI systems than their less-structured, generic counterparts. Consequently, the experimental protocol for all three datasets involved providing the large models with contextual background information on academic scholars and instructing them to perform a binary classification task to determine if an email was fraudulent. To rigorously evaluate the performance boundaries of these models, the Kimi 1000 dataset was selected for stress-testing. Furthermore, to establish a comprehensive benchmark for MLLMs, 200 image-based attack samples were incorporated.

\section{Related Work}

Our work is built upon a deep understanding of MLLMs, a critical analysis of existing benchmarks, and research into the specific applications of AI in cybersecurity.

\subsection{Multimodal Large Language Models (MLLMs)}
With the rapid development of Large Language Models (LLMs), MLLMs have also achieved significant advancements. Many pioneering works have effectively bridged the gap between vision and language by integrating additional modality inputs. For instance, \textbf{BLIP-2}~\citep{li2023blip} encodes images using a Vision Transformer (ViT) and employs a Q-Former to map visual features into the language space. The \textbf{LLaVA} \citep{liu2023llava} series utilizing a simple Multi-Layer Perceptron (MLP) as a connector. Similarly, \textbf{mPLUG-Owl2} \citep{ye2023mplug} employs a modality-adaptive module. These advancements provide the foundation for understanding complex, cross-modal cyberattacks.

\subsection{MLLM Evaluation and Benchmarks}
The rapid growth of MLLMs has spurred an urgent need for their comprehensive evaluation. Early benchmarks focused on tasks like \textbf{Visual Question Answering (VQA)}. Frameworks like \textbf{MMBench} \citep{liu2023mmbench} and \textbf{SEED-Bench} \citep{li2023seed} evaluate capabilities through common-sense questions, while \textbf{MMMU} \citep{yue2023mmmu} uses professional exams.
However, existing security benchmarks have limitations. \textbf{AdvBench} \citep{zou2023universal} focuses on adversarial robustness, while \textbf{SafetyBench} \citep{zhang2023safetybench} covers safety dimensions. These benchmarks predominantly rely on \textbf{static prompt datasets}, failing to simulate dynamic real-world attackers, multilingual prompts, or human-centric factors.

\subsection{AI-driven Threat Landscape and Social Engineering}
AI-driven phishing detection has evolved from feature-based ML to deep learning. However, datasets often fail to reflect personalized \textbf{spear-phishing attacks}. With the proliferation of Vision-Language Models (VLMs), the attack surface has expanded to the multimodal domain (images, QR codes). Understanding implicit meanings requires sophisticated \textbf{Theory of Mind (ToM)} \citep{zhang2025ciibench}. Furthermore, social engineering success is closely related to the target's emotional state \citep{khadka2024survey}.

\subsection{LLM-driven Defense and Jailbreaking}
LLMs are both defensive tools and attack vectors. Attackers use them to generate phishing emails, or target them via \textbf{Jailbreaking}. Techniques include \textbf{weak-to-strong jailbreaking} \citep{zhao2024weak}, \textbf{GPTFuzzer} \citep{gptfuzzer}, \textbf{MathPrompt} \citep{bethany2024mathprompt}, \textbf{xJailbreak} \citep{lee2025xjailbreak}, and \textbf{Distraction-based Attack Prompts} \citep{xiao2024distract}. Frameworks like \textbf{EasyJailbreak} \citep{zhou2024easyjailbreak} organize these methods. AdapT-Bench integrates these simulations to evaluate intrinsic security.
\section{The AdapT-Bench}

\subsection{Overview}
\begin{figure}[ht!]
    \centering
    % NOTE: Ensure '232323.png' exists in your directory or replace with correct filename.
    \includegraphics[width=\columnwidth]{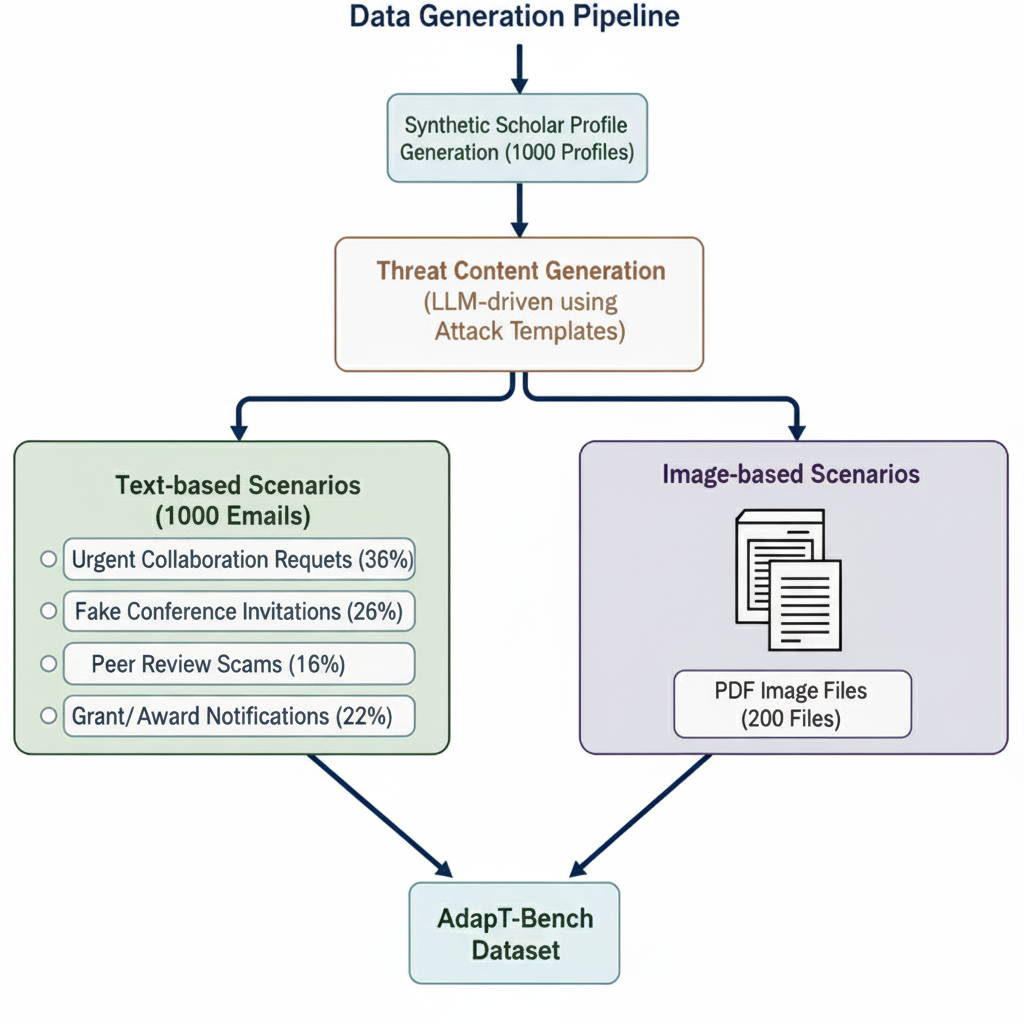}
    \caption{AdapT-Bench, showcasing both text-based and image-based attack scenarios.}
    \label{fig:strategy_distribution}
\end{figure}

AdapT-Bench includes a diverse range of attack vectors, including highly personalized phishing emails and graphic-based phishing attempts rich in contextual information. It assesses if MLLMs can leverage contextual understanding of user profiles and emotional states to defend against targeted threats. The benchmark includes three datasets of varying complexity. Figure~\ref{fig:strategy_distribution} provides representative examples.

\subsection{Data Curation Process}
\subsubsection{Data Collection}
We developed a synthetic academic threat simulation pipeline.
\textbf{Synthetic Scholar Profile Database:} We generated 1000 synthetic scholar profiles reflecting diverse disciplines, career stages, and affiliations. This enables personalized attacks without ethical risks.

\subsubsection{Data Filtering}
We generated data using \texttt{kimi-k2-0711-preview} (not a test model) and implemented a two-stage filtering process for the \textbf{Filtered Clean dataset} ($n=1,054$).

\paragraph{First Stage: Automated Quality Filtering.}
We used a classifier to check content coherence, OSINT integration, and contextual relevance. Only instances scoring above 2 across all dimensions were retained.

\paragraph{Second Stage: Human Verification.}
Cases underwent manual review for Plausibility, Subtlety, and Clarity.

\paragraph{Score Distribution.}
Table~\ref{tab:score_distribution} shows the distribution. Paradoxically, filtering made attacks easier to detect.

\begin{table}[h!]
    \centering
    \setlength{\tabcolsep}{3pt}
    \begin{tabular*}{\columnwidth}{@{\extracolsep{\fill}}l c c l}
        \toprule
        \textbf{Score} & \textbf{Count} & \textbf{Percentage} & \textbf{Status} \\
        \midrule
        2.0 & 10 & 0.95\% & Just Passed \\
        2.2 & 112 & 10.63\% & Good \\
        2.3 & 6 & 0.57\% & Good \\
        2.4 & 853 & 80.93\% & High Quality \\
        2.6 & 72 & 6.83\% & Excellent \\
        2.8 & 1 & 0.09\% & Outstanding \\
        \bottomrule
    \end{tabular*}
    \caption{Score Distribution of the Filtered Dataset.}
    \label{tab:score_distribution}
\end{table}

\subsection{Threat Generation Pipeline}
We used carefully crafted attack templates: \textbf{Grant/Award Notifications}, \textbf{Urgent Collaboration Requests}, \textbf{Peer Review Scams}, and \textbf{Fake Conference Invitations}.

\subsubsection{Strategies Batch 1 Dataset}
\begin{figure}[ht!]
    \centering
    \includegraphics[width=\columnwidth]{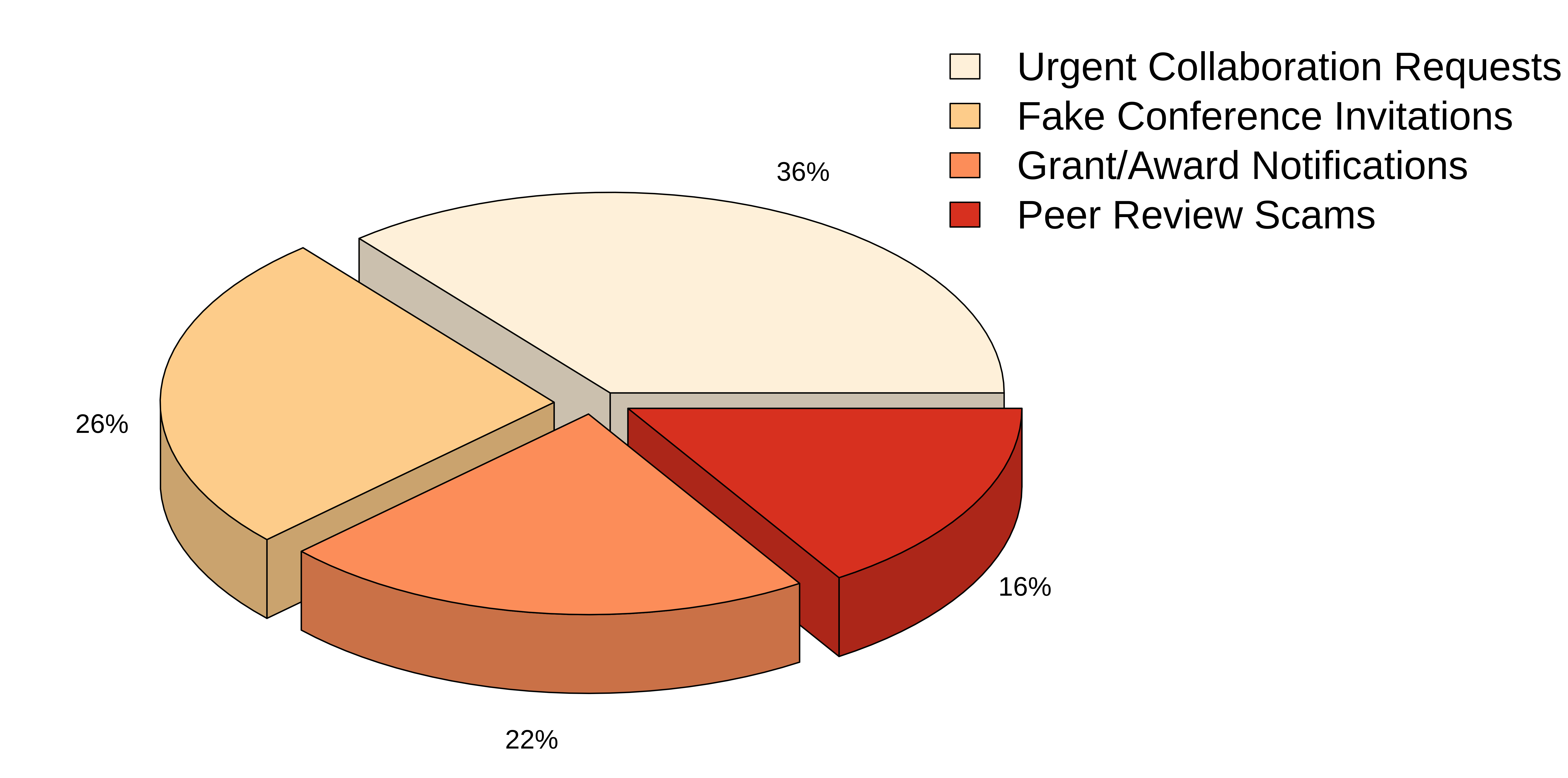}
    \caption{Distribution of the 1000 generated text-based phishing attacks (Strategies Batch 1).}
    \label{fig:dist1000}
\end{figure}

1000 high-quality text-based emails generated from templates. Distribution: Urgent Collaboration (36\%), Fake Conference (26\%), Grant/Award (22\%), Peer Review (16\%). See Figure~\ref{fig:dist1000}.

\subsubsection{Filtered Clean Dataset}
\begin{figure}[ht!]
    \centering
    \includegraphics[width=\columnwidth]{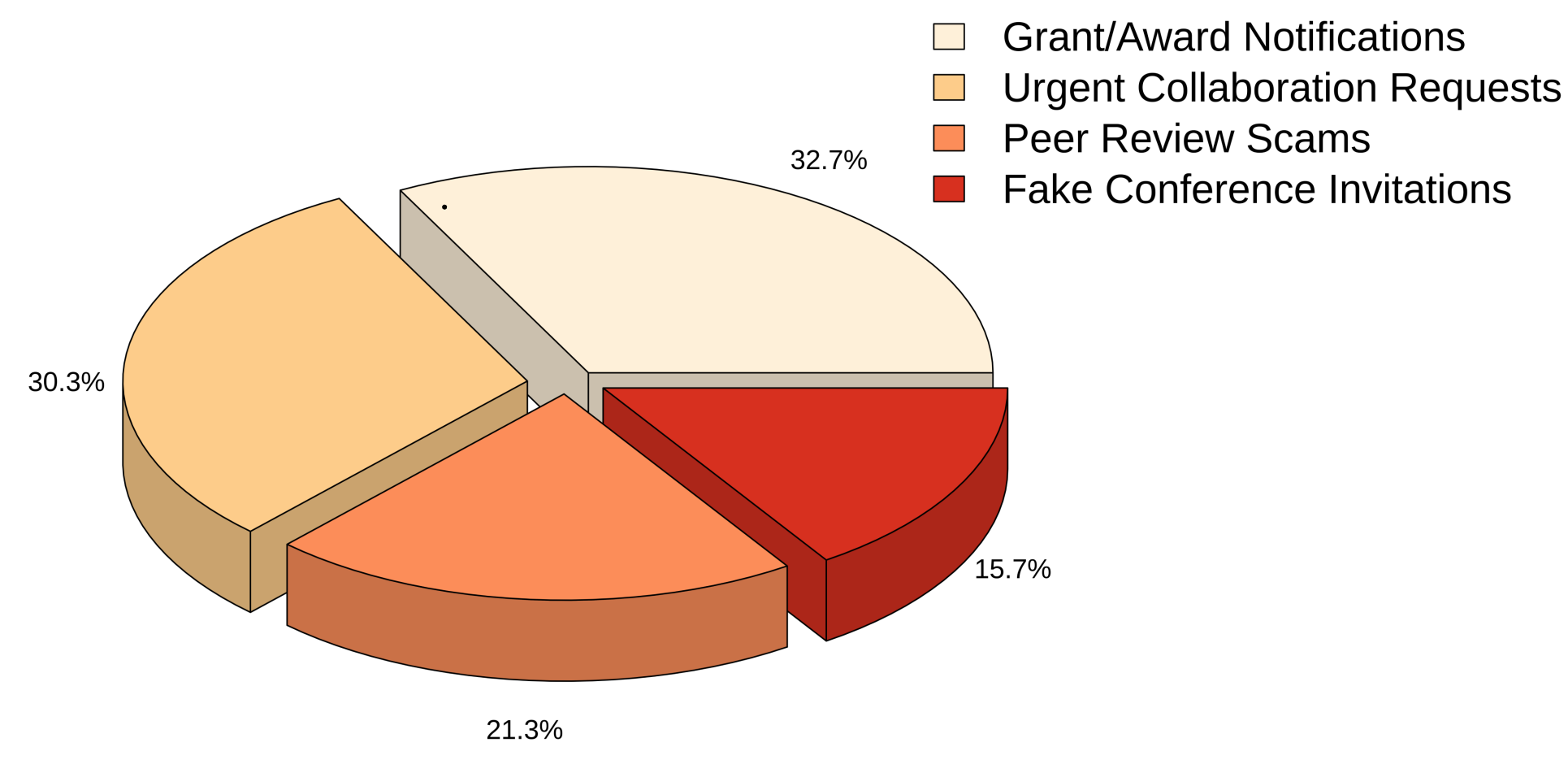}
    \caption{Strategy distribution for the 1,054 instances in the Filtered Clean dataset.}
    \label{fig:dist1054}
\end{figure}

1054 unique instances after rigorous filtering. Distribution: Grant/Award (32.7\%), Urgent Collaboration (30.3\%), Peer Review (21.3\%), Fake Conference (15.7\%). See Figure~\ref{fig:dist1054}.

\subsubsection{Kimi 1000 Dataset}
Generated from non-fixed templates, resulting in random/rougher content. Includes 200 image-based samples.

\section{Experiment}

\subsection{Experimental Setup}
The experiment consists of a baseline performance test (zero-shot) on all three datasets, followed by stress tests (multimodal, emotion, injection) on the most challenging dataset (\textbf{Kimi 1000}).

\paragraph{Models Evaluated:}
\textbf{GPT-4o}, \textbf{GPT-5}, \textbf{Gemini 2.5 Pro}, \textbf{Gemini 2.5 Flash}, \textbf{DeepSeek-V3.2-Exp}, \textbf{Grok 3}, and \textbf{Grok-4-Fast-Reasoning}.

\paragraph{Prompting Strategies:}
\begin{itemize}
    \item \textbf{Neutral Prompt:} Baseline query.
    \item \textbf{Emotional State Prompts:} Simulating states like \textit{anxious}.
    \item \textbf{Cross-Lingual Prompts:} Chinese query for English email.
    \item \textbf{Jailbreak Injection:} Instruction to ignore safety protocols.
\end{itemize}

\subsection{Evaluation Metrics}
\textbf{Accuracy (ACC)} and \textbf{False Negative Rate (FNR)}. FNR is critical as it represents successful attacks.

\section{Results and Analysis}

\subsection{Overall Performance}
Table~\ref{tab:zeroshot_results} summarizes zero-shot capabilities. Most models collapsed on the \textbf{Kimi 1000} dataset. \textbf{Gemini 2.5 Pro} achieved the highest accuracy. \textbf{GPT-4o} showed near-total failure on Kimi 1000 ($FNR=99.8\%$).

\begin{table*}[ht!]
\centering
\begin{tabular}{lcccccc}
\toprule
\multirow{2}{*}{\textbf{Model}} & \multicolumn{2}{c}{\textbf{Filtered Clean}} & \multicolumn{2}{c}{\textbf{Strategies Batch 1}} & \multicolumn{2}{c}{\textbf{Kimi 1000}} \\
\cmidrule(lr){2-3} \cmidrule(lr){4-5} \cmidrule(lr){6-7}
& \textbf{ACC} & \textbf{FNR} & \textbf{ACC} & \textbf{FNR} & \textbf{ACC} & \textbf{FNR} \\
\midrule
GPT-4o & 40.04 & 59.96 & 29.7 & 70.3 & 0.2 & 99.8 \\
GPT-5 & 84.44 & 15.56 & 75.9 & 24.1 & 28.5 & 71.5 \\
Gemini 2.5 Pro & 99.62 & 0.38 & 99.1 & 0.9 & 99.0 & 1.0 \\
Gemini 2.5 Flash & 71.25 & 28.75 & 66.4 & 33.6 & 48.0 & 52.0 \\
DeepSeek-V3.2-Exp & 75.9 & 24.1 & 59.2 & 40.8 & 31.4 & 68.6 \\
Grok 3 & 95.45 & 4.55 & 93.6 & 6.4 & 51.2 & 48.8 \\
Grok-4-Fast & 80.55 & 19.45 & 78.4 & 21.6 & 47.8 & 52.2 \\
\bottomrule
\end{tabular}
\caption{Zero-Shot Performance (\%) on Phishing Detection Datasets.}
\label{tab:zeroshot_results}
\end{table*}

\subsection{Multimodal Performance}
Table~\ref{tab:acc_prompt_comparison} compares performance using English vs. Chinese prompts on Kimi 1000.

\begin{table}[ht!]
\centering
\small
\setlength{\tabcolsep}{2.5pt}
\begin{tabular}{lcccc}
\toprule
\multirow{2}{*}{\textbf{Model}} & \multicolumn{2}{c}{\textbf{Text-based}} & \multicolumn{2}{c}{\textbf{Image-based}} \\
\cmidrule(lr){2-3} \cmidrule(lr){4-5}
& \textbf{Eng} & \textbf{Chi} & \textbf{Eng} & \textbf{Chi} \\
\midrule
GPT-4o & 0.2 & 0.1 & 42.6 & 32.5 \\
GPT-5 & 28.5 & 24.6 & 75.6 & 68.4 \\
Gemini 2.5 Pro & 99.0 & 97.3 & 85.6 & 75.2 \\
Gemini 2.5 Flash & 48.0 & 44.3 & 70.1 & 60.9 \\
Grok 3 & 51.2 & 46.5 & 72.7 & 66.2 \\
\bottomrule
\end{tabular}
\caption{Accuracy (\%) comparison by Prompt Language.}
\label{tab:acc_prompt_comparison}
\end{table}

\subsection{Cross-Lingual and Jailbreak Vulnerabilities}
Table~\ref{tab:cross_lingual_jailbreak} highlights vulnerabilities. GPT-4o was particularly susceptible to jailbreak injections ($FNR=100\%$).

\begin{table}[ht!]
\centering
\small
\begin{tabular}{lcc}
\toprule
\textbf{Model} & \textbf{Cross-Lingual FNR} & \textbf{Jailbreak FNR} \\
\midrule
GPT-4o & 99.8\% & \textbf{100\%} \\
Gemini 2.5 Flash & 52.0\% & 85.6\% \\
Gemini 2.5 Pro & 1.0\% & 21.0\% \\
Grok 3 & 48.8\% & 75.3\% \\
\bottomrule
\end{tabular}
\caption{False Negative Rates under Cross-Lingual and Jailbreak scenarios.}
\label{tab:cross_lingual_jailbreak}
\end{table}

\section{Discussion and Conclusion}

Our findings show that while modern MLLMs are capable, their performance degrades significantly under realistic, unstructured conditions (the "quality filtering paradox"). The "human element," simulated via emotional states and cross-lingual prompts, remains a blind spot. Furthermore, vulnerability to jailbreak injections reveals a tension between helpfulness and safety. We release AdapT-Bench to facilitate research into more robust, context-aware AI safety systems.

\bibliography{reference}

@article{greshake2023prompt,
  title={More than you've asked for: A comprehensive analysis of novel prompt injection threats to application-integrated large language models},
  author={Greshake, Kai and Abdelnabi, Sahar and Mishra, Shailesh and Endres, Christoph and Holz, Thorsten and Fritz, Mario},
  journal={arXiv preprint arXiv:2302.12173},
  year={2023}
}

@inproceedings{zhang2025ciibench,
  title={Can MLLMs Understand the Deep Implication Behind Chinese Images?},
  author={Chenhao Zhang and Xi Feng and Yuelin Bai and Xeron Du and Jinchang Hou and Kaixin Deng and Guangzeng Han and Qinrui Li and Bingli Wang and Jiaheng Liu and Xingwei Qu and Yifei Zhang and Qixuan Zhao and Yiming Liang and Ziqiang Liu and Feiteng Fang and Min Yang and Wenhao Huang and Chenghua Lin and Ge Zhang and Shiwen Ni},
  booktitle={ACL},
  year={2025}
}

@misc{khadka2024survey,
  title={A Survey on the Principles of Persuasion as a Social Engineering Strategy in Phishing}, 
  author={Kalam Khadka and Abu Barkat Ullah and Wanli Ma and Elisa Martinez Marroquin},
  year={2024},
  eprint={2412.18488},
  archivePrefix={arXiv},
  primaryClass={cs.CR}
}

@article{zhang2023safetybench,
  title={SafetyBench: Evaluating the Safety of Large Language Models with Multiple Choice Questions}, 
  author={Zhexin Zhang and Leqi Lei and Lindong Wu and Rui Sun and Yongkang Huang and Chong Long and Xiao Liu and Xuanyu Lei and Jie Tang and Minlie Huang},
  journal={arXiv preprint arXiv:2309.07045},
  year={2023}
}

@misc{zou2023universal,
  title={Universal and Transferable Adversarial Attacks on Aligned Language Models},
  author={Andy Zou and Zifan Wang and J. Zico Kolter and Matt Fredrikson},
  year={2023},
  eprint={2307.15043},
  archivePrefix={arXiv},
  primaryClass={cs.CL}
}

@article{yue2023mmmu,
  title={MMMU: A Massive Multi-discipline Multimodal Understanding and Reasoning Benchmark for Expert AGI},
  author={Xiang Yue and Yuansheng Ni and Kai Zhang and Tianyu Zheng and Ruoqi Liu and Ge Zhang and Samuel Stevens and others},
  journal={arXiv preprint arXiv:2311.16502},
  year={2023}
}

@article{li2023seed,
  title={Seed-bench: Benchmarking multimodal llms with generative comprehension},
  author={Li, Bohao and Wang, Rui and Wang, Guangzhi and Ge, Yuying and Ge, Yixiao and Shan, Ying},
  journal={arXiv preprint arXiv:2307.16125},
  year={2023}
}

@article{liu2023mmbench,
  title={Mmbench: Is your multi-modal model an all-around player?},
  author={Liu, Yuan and Duan, Haodong and Zhang, Yuanhan and Li, Bo and Zhang, Songyang and Zhao, Wangbo and Yuan, Yike and Wang, Jiaqi and He, Conghui and Liu, Ziwei and others},
  journal={arXiv preprint arXiv:2307.06281},
  year={2023}
}

@article{ye2023mplug,
  title={mplug-owl2: Revolutionizing multi-modal large language model with modality collaboration},
  author={Ye, Qinghao and Xu, Haiyang and Ye, Jiabo and Yan, Ming and Liu, Haowei and Qian, Qi and Zhang, Ji and Huang, Fei and Zhou, Jingren},
  journal={arXiv preprint arXiv:2311.04257},
  year={2023}
}

@misc{liu2023llava,
  title={Visual Instruction Tuning}, 
  author={Liu, Haotian and Li, Chunyuan and Wu, Qingyang and Lee, Yong Jae},
  journal={NeurIPS},
  year={2023}
}

@article{li2023blip,
  title={Blip-2: Bootstrapping language-image pre-training with frozen image encoders and large language models},
  author={Li, Junnan and Li, Dongxu and Savarese, Silvio and Hoi, Steven},
  journal={arXiv preprint arXiv:2301.12597},
  year={2023}
}

@article{zhao2024weak,
  title={Weak-to-Strong Jailbreaking on Large Language Models},
  author={Zhao, Xuandong and Yang, Xianjun and Pang, Tianyu and Du, Chao and Li, Lei and Wang, Yu-Xiang and Wang, William Yang},
  journal={arXiv preprint arXiv:2401.17256},
  year={2024}
}

@misc{gptfuzzer,
  title={GPTFUZZER: Red Teaming Large Language Models with Auto-Generated Jailbreak Prompts},
  author={Jiahao Yu and Xingwei Lin and Zheng Yu and Xinyu Xing},
  year={2024},
  eprint={2309.10253},
  archiveprefix = {arXiv},
  primaryclass = {cs.AI},
  url={https://arxiv.org/abs/2309.10253}
}

@misc{bethany2024mathprompt,
  title={Jailbreaking Large Language Models with Symbolic Mathematics}, 
  author={Emet Bethany and Mazal Bethany and Juan Arturo Nolazco Flores and Sumit Kumar Jha and Peyman Najafirad},
  year={2024},
  eprint={2409.11445},
  archivePrefix={arXiv},
  primaryClass={cs.CR},
  url={https://arxiv.org/abs/2409.11445}
}

@misc{lee2025xjailbreak,
  title={xJailbreak: Representation Space Guided Reinforcement Learning for Interpretable LLM Jailbreaking}, 
  author={Sunbowen Lee and Shiwen Ni and Chi Wei and Shuaimin Li and Liyang Fan and Ahmadreza Argha and Hamid Alinejad-Rokny and Ruifeng Xu and Yicheng Gong and Min Yang},
  year={2025},
  eprint={2501.16727},
  archivePrefix={arXiv},
  primaryClass={cs.CL},
  url={https://arxiv.org/abs/2501.16727}
}

@inproceedings{xiao2024distract,
    title={Distract Large Language Models for Automatic Jailbreak Attack},
    author={Zeguan Xiao and Yan Yang and Guanhua Chen and Yun Chen},
    booktitle={Proceedings of the 2024 Conference on Empirical Methods in Natural Language Processing (EMNLP)},
    year={2024}
}

@misc{zhou2024easyjailbreak,
  title={EasyJailbreak: A Unified Framework for Jailbreaking Large Language Models}, 
  author={Weikang Zhou and Xiao Wang and Limao Xiong and Han Xia and Yingshuang Gu and Mingxu Chai and Fukang Zhu and Caishuang Huang and Shihan Dou and Zhiheng Xi and Rui Zheng and Songyang Gao and Yicheng Zou and Hang Yan and Yifan Le and Ruohui Wang and Lijun Li and Jing Shao and Tao Gui and Qi Zhang and Xuanjing Huang},
  year={2024},
  eprint={2403.12171},
  archivePrefix={arXiv},
  primaryClass={cs.CL}
}

\appendix
\end{document}